\documentclass[aps,pra,twocolumnm,reprint,superscriptaddress]{revtex4-1}
\usepackage{amsmath,graphicx,hyperref,xcolor}

\usepackage[utf8]{inputenc}
\usepackage[T1]{fontenc}
\usepackage{natbib}
\usepackage{amsmath,graphicx,color,enumitem,hyperref}
\usepackage{caption}
\usepackage{subcaption}

\newcommand{\figref}[1]{Fig.~\ref{#1}}

\begin{document}
	
\title{Waveguide platform for quantum anticentrifugal force}

\author{Andrzej Gajewski}
\affiliation{Faculty of Physics, Astronomy and Informatics, Nicolaus Copernicus University, Grudziadzka 5, 87-100 Torun, Poland}
\author{Daniel Gustaw}
\affiliation{Faculty of Physics, Astronomy and Informatics, Nicolaus Copernicus University, Grudziadzka 5, 87-100 Torun, Poland}
\author{Roshidah Yusof}
\affiliation{School of Microelectronic Engineering, Universiti Malaysia Perlis, 02000 Arau, Perlis, Malaysia}
\author{Norshamsuri Ali}
\affiliation{School of Microelectronic Engineering, Universiti Malaysia Perlis, 02000 Arau, Perlis, Malaysia}
\author{Karolina Słowik}
\affiliation{Faculty of Physics, Astronomy and Informatics, Nicolaus Copernicus University, Grudziadzka 5, 87-100 Torun, Poland}
\author{Piotr Kolenderski}
\affiliation{Faculty of Physics, Astronomy and Informatics, Nicolaus Copernicus University, Grudziadzka 5, 87-100 Torun, Poland}
\email{Corresponding author: kolenderski@fizyka.umk.pl}

\begin{abstract}
This work is a proposal for an experimental platform to observe quantum fictitious anticentrifugal force. We present an analytical and numerical treatment of a rectangular toroidal dielectric waveguide. Solving the Helmholtz equation we obtain analytical solutions for transverse spatial modes and estimate their number 
as a function of system characteristics. On top of that, the analysis of the structure was extended onto a real material platform, a thin film lithium niobate on an insulator rib waveguide. The framework presented here can be directly applied to analyze the phenomenon of quantum anti-centrifugal force. 
\end{abstract}

\maketitle

Among quantum fictitious forces, the quantum anti-centrifugal force (QAF) is the most intriguing, due to its counter-intuitive characteristics. The quantum anticentrifugal potential appears in systems of cylindrical symmetry \cite{Cirone2001}. Among all possible eigensolutions of the corresponding Schr\"{o}dinger equation the subset with a vanishing angular momentum behaves in a stark contrast to classical counterparts. This means that a particle with small angular momentum described by that solution is attracted towards the center of symmetry rather than repulsed \cite{Dandoloff2011,Dandoloff2014}. 
So far, the analysis of the QAF was performed purely theoretically. However, now experimental capabilities have been developed to fabricate microscopic geometries in which quantum phenomena are manifested. Next, analysis methods are now mature enough to realise all the necessary experimental ingredients such as the phase retrieval algorithm or quantum tomography supported by spatially-resolved single photon detection techniques. All of the above bring experiments on QAF into reach. The most straightforward platform is based on a bent waveguide (BW) structure \cite{Dandoloff2011}.

Spatial mode deformation and losses in optical fibers have been analyzed in a variety of cases \cite{Marcuse1976,Snitzer1961,Hu2009a}, including models assuming toroidal and circular shapes \cite{Janaki1990}, bent slab waveguides \cite{Hiremath2005}, bent optical fibers \cite{Klepavcek2011}, large-area multimode fibers \cite{Smith2012} and diffusion waveguides \cite{Conwell1973, Nalesso2009}. 
This problem is treated by solving the Helmholtz propagation equation with appropriate boundary conditions. The Helmholtz problem can be solved analytically when it is separable with respect to its variables. However, this separability does not hold for all coordinate systems \cite{Morse1953}. 
This necessitates a choice of a suitable coordinate system and boundary conditions that allows one to separate different spatial variables and to solve the eigenmode problem analytically. 
Another problem is related to geometries with sharp corners, e.g. a rectangular one, which require approximations such as the assumption of strong field confinement in the waveguide \cite{Marin2004,Marin2010}. With current fabrication techniques, production of microscaled waveguides has become feasible, and even small bending radii can be achieved. Rectangular shapes are typically obtained in the common bottom-up complementary metal-oxide-semiconductor (CMOS) fabrication scheme \cite{Beals2008}. 
These popular structures have been investigated in the context of spatial mode cross-talk \cite{Gabrielli2012} or chromatic dispersion \cite{Zhang2011}. 

\begin{figure}[htb]
\centering
\includegraphics[width=0.85\columnwidth,keepaspectratio]{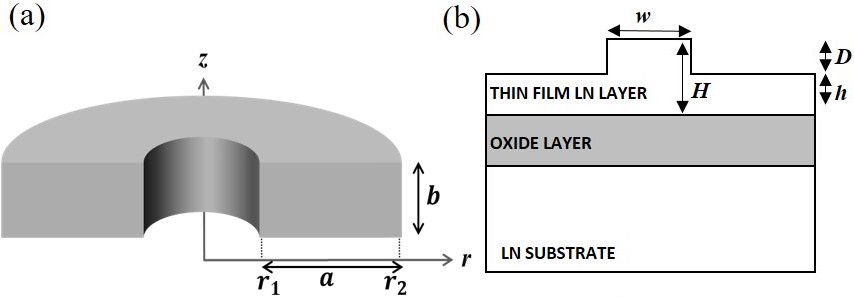}
\caption{\textbf (a) A toroidal waveguide, which is investigated here, has a rectangular profile of dimensions $a$ and $b$. We denote the interior (exterior) bending radius $r_1$ ($r_2$) and the upper (lower) wall position as $z_0$ ($-z_0$), where $z_0 = b/2$. The analysis of the system is simplified in a cylindrical coordinate system.(b)The cross section view of lithium niobate on insulator (LNOI) rib waveguide with $h = 100$ nm, $D = 500$ nm and $W = 1000$ nm.}
\label{fig:setup}
\end{figure}

In this work, we study BW eigenmodes in the context of QAF. 
Our goal it to identify the modes localized near the inner radius of the structure. Any deformation of eigenmodes with respect to their 'straight waveguide' counterparts can be understood in analogy to the phenomenon of a ficticious anticentrifugal force acting on a quantum particle in a cylindrically-symmetric potential characterized with a vanishing angular momentum quantum number \cite{Dandoloff2014}. On the other hand, modes pushed towards the outer radius of the BW are analogons of quantum particle states with larger angular momenta.
In order to find eigenmodes we solve the Helmholtz equation and derive modes expressed by a combination of Bessel functions. Remarkably, the framework developed in the context of radio frequency analysis in Refs. \cite{Cochran1966,Marcatili1969} can be easily adapted into the optical domain. The  analysis supported by perturbative methods allows us to provide an approximate analytical formula for the number of modes as a function of the characteristics of our system. We present and discuss example analytical solutions and successfully verify them against solutions obtained by rigorous numerical simulations of full-vectorial Maxwell's equations. A comparison to mode profiles of a realistic rib fiber structure on a substrate is finally provided.

 First, we analytically investigate the problem of the propagation of an electromagnetic (EM) wave in a rectangular toroidal dielectric as shown in \figref{fig:setup}(a). It is assumed that the index of refraction in the waveguide $n_\mathrm{w}$ is much higher than that in its surroundings $n_\mathrm{s}$. A toroidal waveguide has a rectangular profile with dimensions $a$ and $b$. We assume an interior (exterior) bending radius of $r_1$ ($r_2$). 
 We will
 derive spatial mode profiles of the electric field $\vec E (\vec r) e^ {- i \omega t}$ of an EM wave propagating along the bent structure of the waveguide. Here, $\omega$ is the mode eigenfrequency. We assume the polarization direction of the electric field to be parallel to the $r$-axis during the entire propagation. This is clearly an approximation, which nevertheless yields solutions in very good agreement with the numerical ones, as we will later show. Then, the vectorial notation can be dropped.  
 In order to find $E (\vec r)$ we solve the Helmholtz equation $\Delta E + (k_{w,s}^2 -1/r^2) E=0$, where $k_{w,s}= k n_{w,s} $ is the momentum of the EM field inside (subscript $w$) or outside (subscript $s$) the waveguide. We denote free-space momentum as $ k = 2 \pi/ \lambda$ where $\lambda=\frac{2\pi c}{\omega}$ is the free-space  wavelength of the EM wave. 
In cylindrical coordinates $(r,\phi,z)$ the Helmholtz equation takes the form:
\begin{align}
 \frac{1}{r} \frac{\partial E}{\partial r} +
    \frac{ \partial^2 E}{\partial r^2}+
  \frac{1}{r^2} \frac{\partial^2 E}{\partial \phi^2}+
   \frac{\partial^2 E}{\partial z^2} + (k_{w,s}^2 - \frac{1}{r^2}) E=0.
\label{eq:Helm:1}
\end{align}
This particular choice of coordinate system, under the assumption assumption of strong confinement, permits variable separation $ E(r,\phi,z) = R(r) \Phi(\phi) Z(z)$. Ignoring trivial solutions, we can rewrite \eqref{eq:Helm:1} in the following form:
\begin{align}
     \frac{1}{rR} \frac{\partial R}{\partial r} +
    \frac{1}{R} \frac{ \partial^2 R}{\partial r^2}          +
    \frac{1}{r^2 \Phi} \frac{\partial^2 \Phi}{\partial \phi^2}+
    \frac{1}{Z} \frac{\partial^2 Z}{\partial z^2} + k_{w,s}^2 -\frac{1}{r^2} =0.
    \label{eq:radial}
\end{align}
The independence of variables leads to a system of equations:
\begin{eqnarray}
    \frac{1}{\Phi} \frac{\partial^2 \Phi}{\partial \phi^2} &=& -m^2 \\
    \frac{1}{Z} \frac{\partial^2 Z}{\partial z^2} &=& \begin{cases} -\beta_w^2, \mbox{ if } |z| < z_0 \\ \beta_s^2,  \mbox{ if } |z| > z_0 \end{cases} \label{eq:Z} \\ 
    \frac{1}{r} \frac{\partial R}{\partial r} +\frac{ \partial^2 R}{\partial r^2} +(-\frac{m^2 +1}{r^2} + h_{w,s}^2) R &=& 0, \label{eq:R}
\end{eqnarray}
where $m$, $\beta_{w,s}$ are positive constants originating from the variable separation.   We assume exponential decay outside the waveguide and oscillating solutions inside. These assumptions lead to difference in sign in the right hand side of \eqref{eq:Z}. For the sake of clarity we introduce the following definitions: $ h_w^2 = k_w^2-\beta_w^2$, $h_s^2 = k_s^2+\beta_s^2$. This way, $k_w$ is the total momentum of the EM field inside the waveguide and $\beta_w$ can be interpreted as its projection in the direction of the $Z$-axis. Therefore, $h_w$ is the total momentum in a plane perpendicular to $z$-axis.

Inside the waveguide, $h_w$  is real, resulting in oscillating solutions in the form of a linear combination of Bessel functions along the $r$ axis. Analogously, one obtains oscillating solutions for functions along $z$ and $\phi$:
\begin{subequations}
\begin{eqnarray}
    \Phi_m(\phi) &=& C_{1,m} \sin (m \phi) +  C_{2,m} \cos (m \phi) \label{eq:gensolution:1} \\
    Z_{\beta_w}(z) &=& A_{m} \sin (\beta_w z) + B_{m} cos (\beta_w z) \\
    R_{m \beta_w}(r) &=& \sin(p) J_{\alpha} (h_w r) +  \cos(p) Y_{\alpha}(h_w r)
    \label{eq:gensolution:3}
\end{eqnarray}
\end{subequations}
where  $C_{j,m}, A_m, B_m, p$ are real constants and $\alpha = \sqrt{m^2 +1}$. We  assume no reflection at the end of the waveguide and also consider waves propagating clockwise. 
%%Then our equation \eqref{eq:gensolution:1} simplifies to:  
%%\begin{eqnarray}
%%    \Phi_m(\phi) &=& e^{i m \phi}  \label{eq:solution:1}
%%\end{eqnarray}
%%where we omit unimportant multiplicative factors.
Since we assume the electric field to be $\hat r$ polarized during the propagation, our magnetic field has components along the $z$-axis and along $\hat \phi$. 
Please note that the neglected components scale like $r^{-2}$ and are insignificant in structures of large bending radii relative to the respective wavelength. As we will show, this polarization allows for perfect confinement of the EM wave along the $r$-axis.  In our case, the magnetic field is given by $
\frac{\partial E}{\partial z} \hat \phi - \frac{1}{r}\frac{\partial E}{\partial \phi} \hat z= i \omega \vec B$.
Boundary conditions for magnetic and electric fields force continuity of $\Phi_m(\phi)$, $R_{m \beta}(r)$, $Z_{\beta}(z)$ and $Z'_{\beta}(z)$ on all boundaries.  Continuity of the radial component of the electric and the magnetic field leads us to the conclusion that $h_w = h_s = h$ and as a result, $k^2 n_\mathrm{w}^2 -\beta_w^2 = k^2 n_\mathrm{s}^2 + \beta_s^2$.
The explicit form of the boundary conditions reads:
\begin{subequations}
\begin{eqnarray}
    A_m \sin (\beta_w z_0) + B_{m} cos (\beta_w z_0) &=& D_m e^{ -\beta_s z_0} \label{eq:BoundEZ1}\\
    \beta_w (A_m \cos (\beta_w z_0) - B_m \sin (\beta_w z_0) ) &=& - \beta_s D_m e^{- \beta_s z_0} \label{eq:BoundBZ1}\\
    - A_m \sin (\beta_w  z_0) + B_{m} \cos (\beta_w z_0) &=& E_m e^{- \beta_s z_0} \label{eq:BoundEZ2}\\
    \beta_w (A_m \cos (\beta_w z_0) + B_m \sin (\beta_w z_0) ) &=&  \beta_s E_m e^{ -\beta_s z_0} \label{eq:BoundBZ2}\\
    \sin (p) J_{\alpha} (h r_1) +  cos(p) Y_{\alpha}(h r_1) &=& 0  \label{eq:R:BoundR1}\\
    \sin (p) J_{\alpha} (h r_2) +  \cos (p) Y_{\alpha}(h r_2) &=& 0,  \label{eq:R:BoundR2}
\end{eqnarray}
\end{subequations}
where  Eq.~(\ref{eq:BoundEZ1},\ref{eq:BoundEZ2}) and (\ref{eq:BoundBZ1},\ref{eq:BoundBZ2}) are conditions for the electric and the magnetic fields respectively and $D_m, E_m$ are real constants. For symmetry reasons $D_m = \pm E_m$, solutions corresponding to $D_m = E_m$ ($D_m = - E_m$ ) will be referred to as 'odd' ('even') with the subscript 'o' ('e').

Let us analyze the boundary condition in the z-axis direction. Since the electric field is $\hat r$-polarised, the same radial components of the electric and magnetic fields vanish. This leads to conditions of perfect confinement along the $r$ axis, Eq.~(\ref{eq:R:BoundR1},\ref{eq:R:BoundR2}). The global phase is unimportant and we set $D_m = 1$. It follows that:
\begin{align}
     \tan (\beta_{w,o} z_0) &=  {\sqrt{k^2(n_\mathrm{w}^2  - n_\mathrm{s}^2) - \beta_{w,o}^2}/\beta_{w,e}}, \label{eq:oddsolution}\\ 
     \cot (\beta_{w,e} z_0) &= - {\sqrt{k^2(n_\mathrm{w}^2  - n_\mathrm{s}^2) - \beta_{w,e}^2}/\beta_{w,e}}.
    \label{eq:evensolution}
\end{align} 
The number of roots of the above equations can be estimated as:
\begin{equation}
     N_o^\mathrm{max} = \lceil \frac{k b}{2 \pi} \sqrt{n_\mathrm{w}^2 - n_\mathrm{s}^2}\rceil,\,\, 
     N_e^\mathrm{max} = \lceil \frac{k b}{2 \pi} \sqrt{n_\mathrm{w}^2 - n_\mathrm{s}^2} -\frac{1}{2}\rceil,
    \label{eq:nmax}
\end{equation} 
where $\lceil$ $\rceil$ is the ceiling function. Note that the boundary conditions results in a discrete set of solutions numbered by $N$. From now on we will denote all respective parameters and functions accordingly: $\beta_i$, $h_i$. By $\beta_i$ ($h_i$) we will denote members of sorted set of solutions to both \eqref{eq:oddsolution} and \eqref{eq:evensolution}, where $i < N^\mathrm{max} = N_e^\mathrm{max} + N_o^\mathrm{max}$.

Now we move on to the boundary condition for the electric field amplitude in the $r$-axis direction, which will let us determine the allowed values of the $p$ parameter defined in \eqref{eq:gensolution:3}. The amplitude has to vanish at $r_1$ and $r_2$ resulting in conditions (\ref{eq:R:BoundR1}, \ref{eq:R:BoundR2}). For a given $m$, the condition for existence of non-trivial solutions of \eqref{eq:R:BoundR1} and \eqref{eq:R:BoundR2} has the following form:
\begin{equation}
 0 =J_{\alpha} (h_i r_2) Y_{\alpha}(h_i r_1) - J_{\alpha} (h_i r_1) Y_{\alpha}(h_i r_2), \label{eq:eqsolnum}
\end{equation}
with $ p_{\alpha} = \arctan (-{Y_{\alpha}(h_i r_j)}/{J_{\alpha}(h_i r_j)})$, and $j=1,2$.
 We are going to estimate the number of possible values of $m$ for which the equation above has a solution. A perturbative analysis supplemented by numerical simulations  leads us to a simple formula:
\begin{align}
l_i^\mathrm{max} = \lfloor \sqrt{k^2 n_\mathrm{w}^2 -\beta_\mathrm{i}^2} (r_2-r_1)/\pi \rfloor.
\label{eq:lmax}
\end{align}
Note that this is the upper limit for the mode number. Each of the predicted modes must fulfil an additional condition: $n_\mathrm{s}< n_\mathrm{eff}< n_\mathrm{w}$. To derive \eqref{eq:lmax} we assume that the field propagating along the waveguide does not decay and thus $m$, which might have non-integer values, must be real. The latter point might seem surprising at the first glance: In a real waveguide significant radiation decay is expected due to bending. However, the assumption of negligible losses follows naturally from the previously made assumption of large index ratio $n_\mathrm{w}/n_\mathrm{s}\rightarrow\infty$ and the resulting boundary conditions. Since the field vanishes at the side walls of the waveguide, there is almost no energy dissipation into the surroundings. Alternatively, the same assumption could be understood in terms of total internal reflection at the interface with infinite index contrast.

Next, we present a derivation of formula (\ref{eq:lmax}).  The aim is to find the number of all possible solutions of \eqref{eq:eqsolnum} as a function of $r_1$, $r_2$ and the momentum $h$ in a plane perpendicular to $z$. To solve the Helmholtz equation   ($\ref{eq:R}$), we use a perturbative analysis:  
\begin{align}
  -\frac{ \partial^2 R }{\partial r^2} +\eta \left(\alpha^2 (\frac{1}{r^2}-\frac{1}{\bar{r}^2} ) - \frac{1}{r} \frac{\partial }{\partial r} \right) R  = \left(  h_\mathrm{i}^2-\frac{\alpha^2}{\bar{r}^2} \right) R,
\end{align}
where for simplicity we introduce the following notation: $\bar{r}=(r_2+r_1)/2$, $\Delta r = r_2 - r_1$ and $\eta$ is a perturbative parameter. Setting $\eta=1$ yields \eqref{eq:R}. The zeroth order equation ($\eta=0$):
\begin{align}
- \frac{ \partial^2 R^0(r) }{\partial r^2}   =
   \left[h_\mathrm{i}^2-(\frac{\alpha^0}{\bar{r}})^2\right] R^0(r)
\label{h0}
\end{align}
has the following solution $R^0(r)= \sin[(r-r_1) \sqrt{h_\mathrm{i}^2-(\frac{\alpha^0}{\bar{r}})^2} ]$,
where the argument is chosen to fulfill the boundary condition given by \eqref{eq:R:BoundR1}. The second condition given by \eqref{eq:R:BoundR2} results in the expression for the number of oscillations:	\mbox{${l \pi} = \sqrt{h_\mathrm{i}^2 - ({\alpha^0}/{\bar{r})^2}} \Delta r$.} This leads to the zeroth order relation for the possible values of $m$:
\begin{align}
	m_{\mathrm{i},\mathrm{l}}^0 = \bar{r} \sqrt{h_\mathrm{i}^2-\left(l \pi\right)^2 (\Delta r)^{-2} -\bar{r}^{-2}}.	\label{eq:m0}
\end{align}
This can be used to set an upper bound for the number of modes $l_\mathrm{i}$. The parameter $m$ is real, which means the expression under the square root must be positive. In the context of QAF, this parameter corresponds to the angular momentum of a particle subject to a quantum anticentrifugal potential, attractive for small and repulsive for large $m$. The maximal number of modes $l_\mathrm{i}^\mathrm{max}$ corresponds to the maximal value of $l_\mathrm{i}$ for which $m_{\mathrm{i},\mathrm{l}}^0$ is real:
\begin{align}
	h_\mathrm{i} >  \frac{l_i \pi}{\Delta r}\sqrt{1 + (\frac{\Delta r}{\bar{r} l_i \pi})^2} \approx \frac{l \pi}{\Delta r}.
\end{align}
In consequence, we arrive at \eqref{eq:lmax}. The zeroth-order approximation $l_\mathrm{i}^\mathrm{max}$ is valid everywhere except for very wide structures supporting a large number of radial modes. Including higher perturbation orders only shifts the values of $m$, but the number of solutions $l^{\text{max}}$ remains constant. As we have seen, the finite number of modes originates from the condition given in \eqref{eq:eqsolnum}, which is a function of $m$. This can be understood based on the mathematical behaviour of the Bessel functions: they change their character from oscillating to exponential when the order $m$ exceeds the value of their argument.

\begin{table}[h!]

  \begin{center}
    \begin{tabular}{c|ccc}
    \hline
%    $k$ [ $\mu m^{-1}$ ]& & 12.5664 & \\
%    $n_{max}$ &  & $3$ & \\
%    \hline
   $n$ & $1$ & $2$ & $3$\\
    \hline
    $\beta$  [ $\mu m^{-1}$ ] & 5.03 & 9.94 & 14.46\\
    $h$  [ $\mu m^{-1}$ ] & 17.35 & 15.09  & 1.08\\
   % $n_\mathrm{eff}$ & ? & ?\\
%    $l_{max}$ & 7 & 6 & 4 \\
    \hline
    $(m, n_\mathrm{eff})$ & $i=1$ & $i=2$ & $i=3$ \\
    \hline
    $l=1$ & (20.54, 2.03) &  (17.391, 1.75) & (11.53, 1.22) \\
    $l=2$ & (16.50, 1.86) & (13.54, 1.58) & (8.08, \textcolor{red}{1.04}) \\
    $l=3$ & (13.23, 1.69) & (10.41, 1.40) & (4.56, \textcolor{red}{0.6}) \\
    $l=4$ & (10.26, 1.46) &  (7.15, \textcolor{red}{1.02}) & --- \\
    $l=5$ & (6.03, \textcolor{red}{0.9}) & --- & --- \\ 
    \end{tabular}
  \end{center}
  \caption{The parameters of the spatial modes for the waveguide geometry specified in the main text. The value of $m$ is computed based on \eqref{eq:eqsolnum}. The values of $n_\mathrm{eff} \lesssim 1$ marked in red correspond to non-physical solutions.
  }
  \label{tab:out}
\end{table}

%\begin{figure}[h!]
%\centering
%\includegraphics[width=\columnwidth,keepaspectratio]{Figure8.jpg}
%\caption{$|E_\mathrm{i,l}^\mathrm{y}|^2$ hybrid mode profile of closed straight waveguide at (a)  $i=1,\ l=1$ and (b) $i=2,\ l=2$. }
%\label{fig:straight}
%\end{figure}

%\begin{figure}[h!]
%     \centering
%         \includegraphics[width=0.4\columnwidth]{fig-8a.jpg}
%         \includegraphics[width=0.4\columnwidth]{fig-8b.jpg}
%        \caption{Straight waveguide modes (a)  $(i,l)=(1,1)$, (b) $(2,2)$. }
%\label{fig:straight}
%\end{figure}

We now proceed to discuss example analytical solutions corresponding to a multimode waveguide of small bending radii with $r_1=0.5\, \mu$m, $r_2=1.5\, \mu$m and height $b=0.5\, \mu$m. The refractive indices of the waveguide and the surroundings are set at 2.3 and 1, respectively. The spatial mode parameters are computed in accordance to \eqref{eq:nmax}, \eqref{eq:lmax} at  the free-space wavelength of the field $\lambda=800$ nm and given in Table \ref{tab:out}. %Additionally, the profiles of the straight waveguide modes were obtained by setting the infinite value of radius as the characteristics of the BW are asymptotically converging under this condition. The output of this analysis is shown in \figref{fig:straight} and \figref{fig:modes}. As expected, the straight waveguide supports symmetric mode profiles at fundamental and higher order modes as depicted in \figref{fig:straight}(a) and \figref{fig:straight}(b), respectively.

\begin{figure}[h!]
\centering
\includegraphics[width=0.9\columnwidth,keepaspectratio]{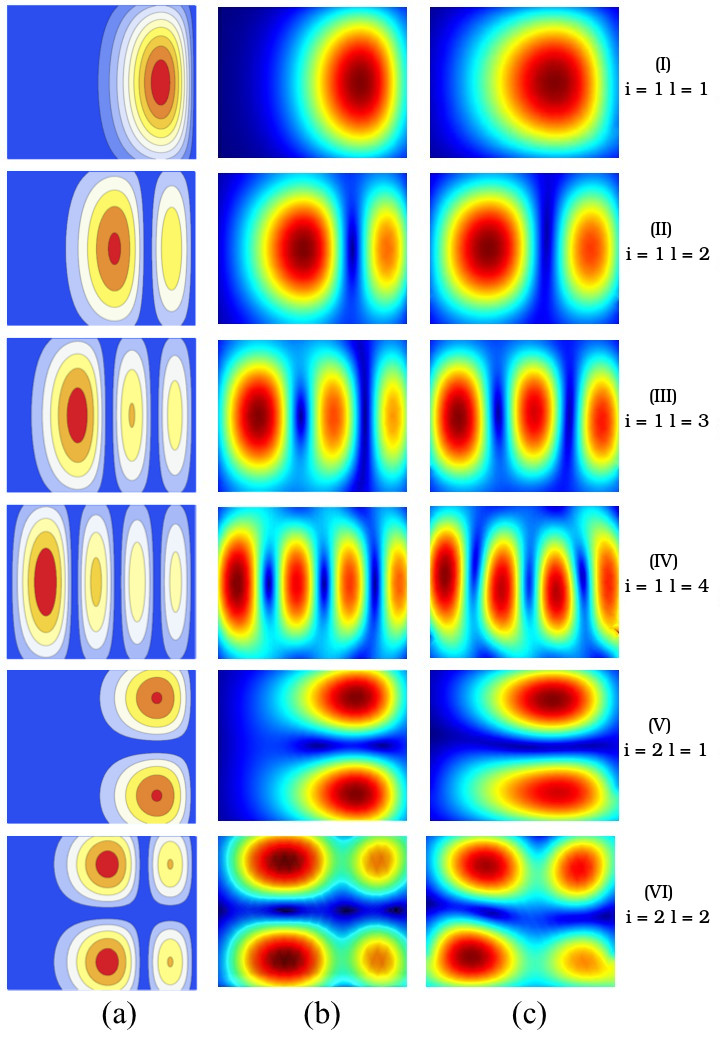}
\caption{Spatial modes of BW in Fig.~\ref{fig:setup}. Analytical (Column a) and numerical (Column b) solutions for the geometry in Fig.~\ref{fig:setup}(a). Modes (Column c) of the LNOI rib waveguide in Fig.~\ref{fig:setup}(b). Rows i-iv (v-vi) correspond to even (odd) solutions. 
}
\label{fig:modes}
\end{figure}

To demonstrate the accuracy and consistency of the spatial profiles given analytically, the same BW was designed and analyzed using the COMSOL software. It was simulated in an EM wave frequency domain by exploiting the finite element method mode solver in cylindrical coordinates. $1\ \mu$m-thick perfectly matched layers and extremely fine meshing of $50$ nm step size were used. Simulations were performed in the module '2.5 D' which assumes axial symmetry, which means we no longer consider a BW but rather a ring with a rectangular cross-section. Nevertheless, the propagation equation for such a ring still has the same form as for the BW. The difference lies in an additional boundary condition for the $\Phi$ function which leads to the requirement that $m$ be an integer. In \figref{fig:modes}(a,b) one can see the good agreement between the analytical and numerical solutions, which justifies the analytical model. However, not all modes found analytically could be recreated in simulation. Mode $(i,l)=(2,3)$ could be found only in the case of the rib waveguide. 
%However, when it comes to the number of modes $N^\mathrm{max}$, the analytical model allows for a higher number of supported solutions, up to $(i,l)=(2,3)$ and $(3,1)$ as stated in Tab.~\ref{tab:my_label}. 
The BW mode profiles are also distorted and shifted towards the sidewalls of the BW as shown in \figref{fig:modes}(a,b). The shift is larger for a reduced bending radius. For \mbox{$r_2 =1$ $\mu \mathrm{m}$} the fundamental mode ($i=l=1$) is shifted significantly towards the outer wall. As $l$ grows, modes shift toward the inner sidewall, which is depicted in \figref{fig:modes}ii-vi. To quantify these observations we introduce the mode average radial position $r_\text{av}(i,l)=\int\int |E_{i,l}(r, 0,z)|^2  r \text{d}r \text{d}z/\int\int |E_{i,l}(r, 0,z)|^2 \text{d}r \text{d}z$.
 This quantity is computed for each of the modes in Tab.~\ref{tab:out} and the analytical and numerical results are compared in Tab.~\ref{tab:my_label}. For the straight waveguide, the average radial position is always in the center of the waveguide \mbox{$r_\text{av}= 1\ \mu\mathrm{m}$}. On the contrary, $r_\text{av}(i,l)$ decreases with increasing mode numbers $i$ and $l$. This counterintuitive behavior was described in a quantum-optical picture through the notion of QAF\cite{Dandoloff2014,Dandoloff2011,Cirone2001}, where in cylindrical geometries, depending on the mode number $l$, the photons can be directed towards or outwards from the radial axis.

\begin{table}[]
    \centering
    \scalebox{0.7}{
    \begin{tabular}{cccccccccc}
   \hline
    Mode & \multicolumn{7}{c}{Average Radial Position $r_\mathrm{av}$ ($\mu m$)}\\
 \cline{2-10}

    Number & \multicolumn{3}{c}{Analytical (Toroid)} & \multicolumn{3}{c}{Numerical (Toroid)}& \multicolumn{3}{c}{Numerical (Rib)}\\

   \hline
   &$i=1$&$i=2$&$i=3$&$i=1$&$i=2$&$i=3$&$i=1$&$i=2$&$i=3$\\
    $l=1$ & 1.29 & 1.27 & 1.21 & 1.21 & 1.17 & 1.05 & 1.17 & 1.12 & 1.00\\
    $l=2$ & 1.13 & 1.09 & \textcolor{red}{0.99} & 1.04 & 0.98 &  - & 0.98 & 0.99 & - \\
    $l=3$ & 1.00 & 0.95 & \textcolor{red}{0.9} & 0.95 &  -   &  - & 0.91 & 1.10 & - \\
    $l=4$ & 0.90 & \textcolor{red}{0.89} & - & 0.95 &  -   &  - & 0.90 &- &
          \end{tabular}
          }
    \caption{Average radial position of waveguide modes computed analytically and numerically. The values marked in red correspond to non-physical solutions (See Tab.~\ref{tab:out}).}
    \label{tab:my_label}
\end{table}

We also extend our analysis to a realistic material platform. We consider a lithium niobate (LN) on insulator (LNOI) due to its production feasibility with low losses within a broad transparency window spanning from $0.5 - 5\ \mu$m. We have simulated a thin film LNOI rib waveguide structure deposited on top of $2\ \mu$m of buried oxide of $500\ \mu$m thick LN substrate as shown in \figref{fig:setup}b, with bending radius, $r=10\ \mu$m. The high index contrast between the LN core ($n_\text{LN}=2.14$) and a glass cladding ($n_\text{SiO2}=1.44$) causes strong confinement across the structure. The eigenmode characteristics are given by the effective refractive index $n_\text{eff} = m/(r_\mathrm{av} k)$. Naturally, a propagation mode is obtained when $n_\text{SiO2}$ < $n_\text{eff}$ < $n_\text{LN}$ condition is fulfilled. The maximum value of $n_\text{eff}$  corresponds to the fundamental eigenmode solution of the Maxwell’s equation. \figref{fig:modes}(c) depicts spatial profiles of rib waveguide modes. Despite the significantly richer, more realistic and asymmetric waveguide geometry, the modes depicted in Fig.~\ref{fig:modes}(c) show striking similarity to those in Fig.~\ref{fig:modes}(a,b). In particular, our analytical model correctly predicts all qualitative features including the modal distortion towards/outwards from the center. The slight asymmetry in the $z$ direction is a natural consequence of the symmetry breaking due to the presence of the substrate (oxide layer, see Fig.~\ref{fig:setup}(b)). 

%\section{Conclusions}
In summary, we have developed an analytical model for spatial modes in a bent rectangular waveguide with an assumption that the index of refraction of the medium is much higher than $1$, thus allowing us to include mode leakage. 
We have juxtaposed that model with accurate numerical solutions for the same geometry, as well as with a feasible experimental platform. The model allows one to predict the spatial mode profiles with a high level of accuracy, as well as estimate their number. Additionally, we have found that the eigenmodes of the investigated structures show an intriguing behavior with their distortion being linked to the mode order $l$ and the curvature of the geometry. 
Please note that similar equations describe the phenomenon of QAF. This is a consequence of the form of Helmholtz equation which in this case is mathematically identical to the Schr\"{o}dinger equation. Thus, the setups proposed in this work may provide a classical platform to test quantum phenomena in curved geometries \cite{Dandoloff2014}.

\medskip
\noindent\textbf{Funding} National Science Centre, Poland, Sonata 12 grant no.~2016/23/D/ST2/02064); Foundation for Polish Science, project First Team co-financed by the European Union under the European Regional Development Fund and project HEIMaT No. Homing/2016-1/8; PROM Project, funded by Polish National Agency for Academic Exchange "NAWA", PPI/PRO/2018/1/00016/U/001; NCU TAPS 2019.

\medskip
\noindent\textbf{Disclosures.} The authors declare no conflicts of interest.

\bibliography{biblio}

\begin{thebibliography}{20}%
\makeatletter
\providecommand \@ifxundefined [1]{%
 \@ifx{#1\undefined}
}%
\providecommand \@ifnum [1]{%
 \ifnum #1\expandafter \@firstoftwo
 \else \expandafter \@secondoftwo
 \fi
}%
\providecommand \@ifx [1]{%
 \ifx #1\expandafter \@firstoftwo
 \else \expandafter \@secondoftwo
 \fi
}%
\providecommand \natexlab [1]{#1}%
\providecommand \enquote  [1]{``#1''}%
\providecommand \bibnamefont  [1]{#1}%
\providecommand \bibfnamefont [1]{#1}%
\providecommand \citenamefont [1]{#1}%
\providecommand \href@noop [0]{\@secondoftwo}%
\providecommand \href [0]{\begingroup \@sanitize@url \@href}%
\providecommand \@href[1]{\@@startlink{#1}\@@href}%
\providecommand \@@href[1]{\endgroup#1\@@endlink}%
\providecommand \@sanitize@url [0]{\catcode `\\12\catcode `\$12\catcode
  `\&12\catcode `\#12\catcode `\^12\catcode `\_12\catcode `\%12\relax}%
\providecommand \@@startlink[1]{}%
\providecommand \@@endlink[0]{}%
\providecommand \url  [0]{\begingroup\@sanitize@url \@url }%
\providecommand \@url [1]{\endgroup\@href {#1}{\urlprefix }}%
\providecommand \urlprefix  [0]{URL }%
\providecommand \Eprint [0]{\href }%
\providecommand \doibase [0]{http://dx.doi.org/}%
\providecommand \selectlanguage [0]{\@gobble}%
\providecommand \bibinfo  [0]{\@secondoftwo}%
\providecommand \bibfield  [0]{\@secondoftwo}%
\providecommand \translation [1]{[#1]}%
\providecommand \BibitemOpen [0]{}%
\providecommand \bibitemStop [0]{}%
\providecommand \bibitemNoStop [0]{.\EOS\space}%
\providecommand \EOS [0]{\spacefactor3000\relax}%
\providecommand \BibitemShut  [1]{\csname bibitem#1\endcsname}%
\let\auto@bib@innerbib\@empty
%</preamble>
\bibitem [{\citenamefont {Cirone}\ \emph {et~al.}(2001)\citenamefont {Cirone},
  \citenamefont {{Rza{\.z}ewski}}, \citenamefont {{Schleich}}, \citenamefont
  {{Straub}},\ and\ \citenamefont {{Wheeler}}}]{Cirone2001}%
  \BibitemOpen
  \bibfield  {author} {\bibinfo {author} {\bibfnamefont {M.~A.}\ \bibnamefont
  {Cirone}}, \bibinfo {author} {\bibfnamefont {K.}~\bibnamefont
  {{Rza{\.z}ewski}}}, \bibinfo {author} {\bibfnamefont {W.~P.}\ \bibnamefont
  {{Schleich}}}, \bibinfo {author} {\bibfnamefont {F.}~\bibnamefont
  {{Straub}}}, \ and\ \bibinfo {author} {\bibfnamefont {J.~A.}\ \bibnamefont
  {{Wheeler}}},\ }\href {\doibase 10.1103/PhysRevA.65.022101} {\bibfield
  {journal} {\bibinfo  {journal} {Phys. Rev. A}\ }\textbf {\bibinfo {volume}
  {65}},\ \bibinfo {pages} {022101} (\bibinfo {year} {2001})},\ \Eprint
  {http://arxiv.org/abs/arXiv:quant-ph/0108069} {arXiv:quant-ph/0108069}
  \BibitemShut {NoStop}%
\bibitem [{\citenamefont {Dandoloff}\ and\ \citenamefont
  {Atanasov}(2011)}]{Dandoloff2011}%
  \BibitemOpen
  \bibfield  {author} {\bibinfo {author} {\bibfnamefont {R.}~\bibnamefont
  {Dandoloff}}\ and\ \bibinfo {author} {\bibfnamefont {V.}~\bibnamefont
  {Atanasov}},\ }\href {\doibase 10.1002/andp.201100136} {\bibfield  {journal}
  {\bibinfo  {journal} {Annalen der Physik}\ }\textbf {\bibinfo {volume}
  {523}},\ \bibinfo {pages} {925} (\bibinfo {year} {2011})}\BibitemShut
  {NoStop}%
\bibitem [{\citenamefont {Dandoloff}\ \emph {et~al.}(2014)\citenamefont
  {Dandoloff}, \citenamefont {{Jensen}},\ and\ \citenamefont
  {{Saxena}}}]{Dandoloff2014}%
  \BibitemOpen
  \bibfield  {author} {\bibinfo {author} {\bibfnamefont {R.}~\bibnamefont
  {Dandoloff}}, \bibinfo {author} {\bibfnamefont {B.}~\bibnamefont {{Jensen}}},
  \ and\ \bibinfo {author} {\bibfnamefont {A.}~\bibnamefont {{Saxena}}},\
  }\href {\doibase 10.1016/j.physleta.2013.12.016} {\bibfield  {journal}
  {\bibinfo  {journal} {Phys. Lett. A}\ }\textbf {\bibinfo {volume} {378}},\
  \bibinfo {pages} {510} (\bibinfo {year} {2014})}\BibitemShut {NoStop}%
\bibitem [{\citenamefont {Marcuse}(1976)}]{Marcuse1976}%
  \BibitemOpen
  \bibfield  {author} {\bibinfo {author} {\bibfnamefont {D.}~\bibnamefont
  {Marcuse}},\ }\href {\doibase 10.1364/JOSA.66.000311} {\bibfield  {journal}
  {\bibinfo  {journal} {J. Opt. Soc. Am.}\ }\textbf {\bibinfo {volume} {66}},\
  \bibinfo {pages} {311} (\bibinfo {year} {1976})}\BibitemShut {NoStop}%
\bibitem [{\citenamefont {Snitzer}(1961)}]{Snitzer1961}%
  \BibitemOpen
  \bibfield  {author} {\bibinfo {author} {\bibfnamefont {E.}~\bibnamefont
  {Snitzer}},\ }\href
  {http://www.opticsinfobase.org/abstract.cfm?URI=josa-51-5-491} {\bibfield
  {journal} {\bibinfo  {journal} {J. Opt. Soc. Am.}\ }\textbf {\bibinfo
  {volume} {51}},\ \bibinfo {pages} {491} (\bibinfo {year} {1961})}\BibitemShut
  {NoStop}%
\bibitem [{\citenamefont {Hu}\ and\ \citenamefont {Menyuk}(2009)}]{Hu2009a}%
  \BibitemOpen
  \bibfield  {author} {\bibinfo {author} {\bibfnamefont {J.}~\bibnamefont
  {Hu}}\ and\ \bibinfo {author} {\bibfnamefont {C.~R.}\ \bibnamefont
  {Menyuk}},\ }\href {\doibase 10.1364/AOP.1.000058} {\bibfield  {journal}
  {\bibinfo  {journal} {Adv. Opt. Photon.}\ }\textbf {\bibinfo {volume} {1}},\
  \bibinfo {pages} {58} (\bibinfo {year} {2009})}\BibitemShut {NoStop}%
\bibitem [{\citenamefont {Janaki}\ and\ \citenamefont
  {Dasgupta}(1990)}]{Janaki1990}%
  \BibitemOpen
  \bibfield  {author} {\bibinfo {author} {\bibfnamefont {M.~S.}\ \bibnamefont
  {Janaki}}\ and\ \bibinfo {author} {\bibfnamefont {B.}~\bibnamefont
  {Dasgupta}},\ }\href@noop {} {\bibfield  {journal} {\bibinfo  {journal} {IEEE
  T. Plasma. Sci.}\ }\textbf {\bibinfo {volume} {18}},\ \bibinfo {pages} {78}
  (\bibinfo {year} {1990})}\BibitemShut {NoStop}%
\bibitem [{\citenamefont {Hiremath}\ \emph {et~al.}(2005)\citenamefont
  {Hiremath}, \citenamefont {Hammer}, \citenamefont {Stoffer}, \citenamefont
  {Prkna},\ and\ \citenamefont {J.}}]{Hiremath2005}%
  \BibitemOpen
  \bibfield  {author} {\bibinfo {author} {\bibfnamefont {K.}~\bibnamefont
  {Hiremath}}, \bibinfo {author} {\bibfnamefont {M.}~\bibnamefont {Hammer}},
  \bibinfo {author} {\bibfnamefont {R.}~\bibnamefont {Stoffer}}, \bibinfo
  {author} {\bibfnamefont {L.}~\bibnamefont {Prkna}}, \ and\ \bibinfo {author}
  {\bibfnamefont {C.}~\bibnamefont {J.}},\ }\href {\doibase
  10.1007/s11082-005-1118-3} {\bibfield  {journal} {\bibinfo  {journal} {Opt.
  Quantum. Electron.}\ }\textbf {\bibinfo {volume} {37}},\ \bibinfo {pages}
  {37} (\bibinfo {year} {2005})}\BibitemShut {NoStop}%
\bibitem [{\citenamefont {Klep{\'a}{\v c}ek}\ and\ \citenamefont
  {{Kalvoda}}(2011)}]{Klepavcek2011}%
  \BibitemOpen
  \bibfield  {author} {\bibinfo {author} {\bibfnamefont {R.}~\bibnamefont
  {Klep{\'a}{\v c}ek}}\ and\ \bibinfo {author} {\bibfnamefont {L.}~\bibnamefont
  {{Kalvoda}}},\ }in\ \href {\doibase 10.1117/12.886869} {\emph {\bibinfo
  {booktitle} {Society of Photo-Optical Instrumentation Engineers (SPIE)
  Conference Series}}},\ \bibinfo {series} {Society of Photo-Optical
  Instrumentation Engineers (SPIE) Conference Series}, Vol.\ \bibinfo {volume}
  {8073}\ (\bibinfo {year} {2011})\BibitemShut {NoStop}%
\bibitem [{\citenamefont {Smith}\ \emph {et~al.}(2012)\citenamefont {Smith},
  \citenamefont {{Sarangan}}, \citenamefont {{Jiang}},\ and\ \citenamefont
  {{Marciante}}}]{Smith2012}%
  \BibitemOpen
  \bibfield  {author} {\bibinfo {author} {\bibfnamefont {R.~C.~G.}\
  \bibnamefont {Smith}}, \bibinfo {author} {\bibfnamefont {A.~M.}\ \bibnamefont
  {{Sarangan}}}, \bibinfo {author} {\bibfnamefont {Z.}~\bibnamefont {{Jiang}}},
  \ and\ \bibinfo {author} {\bibfnamefont {J.~R.}\ \bibnamefont
  {{Marciante}}},\ }\href {\doibase 10.1364/OE.20.004436} {\bibfield  {journal}
  {\bibinfo  {journal} {Opt. Express}\ }\textbf {\bibinfo {volume} {20}},\
  \bibinfo {pages} {4436} (\bibinfo {year} {2012})}\BibitemShut {NoStop}%
\bibitem [{\citenamefont {Conwell}(1973)}]{Conwell1973}%
  \BibitemOpen
  \bibfield  {author} {\bibinfo {author} {\bibfnamefont {E.~M.}\ \bibnamefont
  {Conwell}},\ }\href {\doibase 10.1063/1.1654906} {\bibfield  {journal}
  {\bibinfo  {journal} {Appl. Phys. Lett.}\ }\textbf {\bibinfo {volume} {23}},\
  \bibinfo {pages} {328} (\bibinfo {year} {1973})}\BibitemShut {NoStop}%
\bibitem [{\citenamefont {Nalesso}\ and\ \citenamefont
  {Pigozzo}(2009)}]{Nalesso2009}%
  \BibitemOpen
  \bibfield  {author} {\bibinfo {author} {\bibfnamefont {G.}~\bibnamefont
  {Nalesso}}\ and\ \bibinfo {author} {\bibfnamefont {F.~M.}\ \bibnamefont
  {Pigozzo}},\ }\href {\doibase DOI: 10.1016/j.optcom.2009.05.070} {\bibfield
  {journal} {\bibinfo  {journal} {Opt Commun}\ }\textbf {\bibinfo {volume}
  {282}},\ \bibinfo {pages} {3596 } (\bibinfo {year} {2009})}\BibitemShut
  {NoStop}%
\bibitem [{\citenamefont {Morse}\ and\ \citenamefont
  {Feshbach}(1953)}]{Morse1953}%
  \BibitemOpen
  \bibfield  {author} {\bibinfo {author} {\bibfnamefont {P.~M.}\ \bibnamefont
  {Morse}}\ and\ \bibinfo {author} {\bibfnamefont {H.}~\bibnamefont
  {Feshbach}},\ }\href@noop {} {\emph {\bibinfo {title} {Methods of Theoretical
  Physics}}}\ (\bibinfo  {publisher} {McGraw-Hill Science/Engineering/Math},\
  \bibinfo {year} {1953})\BibitemShut {NoStop}%
\bibitem [{\citenamefont {Marin}\ \emph {et~al.}(2004)\citenamefont {Marin},
  \citenamefont {Lesnic},\ and\ \citenamefont {Mantic}}]{Marin2004}%
  \BibitemOpen
  \bibfield  {author} {\bibinfo {author} {\bibfnamefont {L.}~\bibnamefont
  {Marin}}, \bibinfo {author} {\bibfnamefont {D.}~\bibnamefont {Lesnic}}, \
  and\ \bibinfo {author} {\bibfnamefont {V.}~\bibnamefont {Mantic}},\ }\href
  {\doibase 10.1016/j.jsv.2003.09.059} {\bibfield  {journal} {\bibinfo
  {journal} {Journal of Sound and Vibration}\ }\textbf {\bibinfo {volume}
  {278}},\ \bibinfo {pages} {39} (\bibinfo {year} {2004})}\BibitemShut
  {NoStop}%
\bibitem [{\citenamefont {Marin}(2010)}]{Marin2010}%
  \BibitemOpen
  \bibfield  {author} {\bibinfo {author} {\bibfnamefont {L.}~\bibnamefont
  {Marin}},\ }\href {\doibase 10.1016/j.apm.2009.09.009} {\bibfield  {journal}
  {\bibinfo  {journal} {Applied Mathematical Modelling}\ }\textbf {\bibinfo
  {volume} {34}},\ \bibinfo {pages} {1615} (\bibinfo {year}
  {2010})}\BibitemShut {NoStop}%
\bibitem [{\citenamefont {et. al.}(2008)}]{Beals2008}%
  \BibitemOpen
  \bibfield  {author} {\bibinfo {author} {\bibfnamefont {M.~B.}\ \bibnamefont
  {et. al.}},\ }in\ \href {\doibase 10.1117/12.774576} {\emph {\bibinfo
  {booktitle} {Silicon Photonics III}}},\ Vol.\ \bibinfo {volume} {6898},\
  \bibinfo {editor} {edited by\ \bibinfo {editor} {\bibfnamefont {J.~A.}\
  \bibnamefont {Kubby}}\ and\ \bibinfo {editor} {\bibfnamefont {G.~T.}\
  \bibnamefont {Reed}}},\ \bibinfo {organization} {International Society for
  Optics and Photonics}\ (\bibinfo  {publisher} {SPIE},\ \bibinfo {year}
  {2008})\ pp.\ \bibinfo {pages} {31 -- 44}\BibitemShut {NoStop}%
\bibitem [{\citenamefont {Gabrielli}\ \emph {et~al.}(2012)\citenamefont
  {Gabrielli}, \citenamefont {Liu}, \citenamefont {Johnson},\ and\
  \citenamefont {Lipson}}]{Gabrielli2012}%
  \BibitemOpen
  \bibfield  {author} {\bibinfo {author} {\bibfnamefont {L.~H.}\ \bibnamefont
  {Gabrielli}}, \bibinfo {author} {\bibfnamefont {D.}~\bibnamefont {Liu}},
  \bibinfo {author} {\bibfnamefont {S.~G.}\ \bibnamefont {Johnson}}, \ and\
  \bibinfo {author} {\bibfnamefont {M.}~\bibnamefont {Lipson}},\ }\href@noop {}
  {\bibfield  {journal} {\bibinfo  {journal} {Nature Communications}\ }\textbf
  {\bibinfo {volume} {3}},\ \bibinfo {pages} {1} (\bibinfo {year}
  {2012})}\BibitemShut {NoStop}%
\bibitem [{\citenamefont {Zhang}\ \emph {et~al.}(2011)\citenamefont {Zhang},
  \citenamefont {{Zhou}}, \citenamefont {{Wang}}, \citenamefont {{Huang}},\
  and\ \citenamefont {{Peng}}}]{Zhang2011}%
  \BibitemOpen
  \bibfield  {author} {\bibinfo {author} {\bibfnamefont {W.}~\bibnamefont
  {Zhang}}, \bibinfo {author} {\bibfnamefont {Q.}~\bibnamefont {{Zhou}}},
  \bibinfo {author} {\bibfnamefont {P.}~\bibnamefont {{Wang}}}, \bibinfo
  {author} {\bibfnamefont {Y.}~\bibnamefont {{Huang}}}, \ and\ \bibinfo
  {author} {\bibfnamefont {J.}~\bibnamefont {{Peng}}},\ }in\ \href {\doibase
  10.1117/12.902030} {\emph {\bibinfo {booktitle} {Society of Photo-Optical
  Instrumentation Engineers (SPIE) Conference Series}}},\ \bibinfo {series}
  {Society of Photo-Optical Instrumentation Engineers (SPIE) Conference
  Series}, Vol.\ \bibinfo {volume} {8011}\ (\bibinfo {year} {2011})\BibitemShut
  {NoStop}%
\bibitem [{\citenamefont {Cochran}\ and\ \citenamefont
  {Pecina}(1966)}]{Cochran1966}%
  \BibitemOpen
  \bibfield  {author} {\bibinfo {author} {\bibfnamefont {J.~A.}\ \bibnamefont
  {Cochran}}\ and\ \bibinfo {author} {\bibfnamefont {R.~G.}\ \bibnamefont
  {Pecina}},\ }\href@noop {} {\bibfield  {journal} {\bibinfo  {journal} {Radio
  Science}\ }\textbf {\bibinfo {volume} {1}},\ \bibinfo {pages} {679} (\bibinfo
  {year} {1966})}\BibitemShut {NoStop}%
\bibitem [{\citenamefont {Marcatili}(1969)}]{Marcatili1969}%
  \BibitemOpen
  \bibfield  {author} {\bibinfo {author} {\bibfnamefont {E.}~\bibnamefont
  {Marcatili}},\ }\href@noop {} {\bibfield  {journal} {\bibinfo  {journal}
  {Bell System Technical Journal}\ }\textbf {\bibinfo {volume} {48}},\ \bibinfo
  {pages} {2103} (\bibinfo {year} {1969})}\BibitemShut {NoStop}%
\end{thebibliography}%

\end{document}